\documentclass[twocolumn,pre,superscriptaddress,floatfix,amsmath,amssymb,aps]{revtex4-2}
\pdfoutput=1

\usepackage[utf8]{inputenc}
\usepackage{natbib}
\usepackage{array}
\usepackage{booktabs}
\usepackage{graphicx}
\usepackage{paralist}
\usepackage{color,soul}
\usepackage[colorlinks=true, allcolors=blue]{hyperref}
\usepackage{mathtools}
\usepackage{amsmath}
\usepackage{dcolumn}
\usepackage{amssymb}
\usepackage{gensymb}
\usepackage{mathdesign}
\usepackage{siunitx}
\usepackage{accents}
\usepackage{bm}
\usepackage{dcolumn}

\newcommand{\Monti}[1]{{{\color{black}#1}}}
\newcommand{\Silbert}[1]{{{\color{black}#1}}}

\begin{document}

\title{Emergence of Intermediate Range Order in Jammed Packings}
\author{Joseph M. Monti}
\affiliation{Sandia National Laboratories, Albuquerque, NM 87185, USA}
\author{Ishan Srivastava}
\affiliation{Center for Computational Sciences and Engineering, Lawrence Berkeley National Laboratory, Berkeley, California 94720, USA}
\author{Leonardo E. Silbert}
\affiliation{School of Math, Science, and Engineering, Central New Mexico Community College, Albuquerque, New Mexico 87106, USA}
\author{A. P. Santos}
\affiliation{AMA Inc., Thermal Protection Materials Branch, NASA Ames Research Center, Moffett Field, CA 94035, USA}
\author{Joel T. Clemmer}
\affiliation{Sandia National Laboratories, Albuquerque, NM 87185, USA}
\author{Jeremy B. Lechman}
\affiliation{Sandia National Laboratories, Albuquerque, NM 87185, USA}
\author{Gary S. Grest}
\affiliation{Sandia National Laboratories, Albuquerque, NM 87185, USA}
\date{\today}

\begin{abstract}

We perform a structural analysis of large scale jammed packings of monodisperse, frictionless and frictional spheres to elucidate structural signatures of the static structure factor in the low-to-intermediate wavenumber ($q$) region. 
We employ discrete element method simulations containing up to eighty million particles, in which the particle friction coefficient(s), including sliding, rolling, and twisting interactions, are varied.
At intermediate $q$ values, corresponding to length scales that lie between that of the nearest neighbor primary peak and the system size, we find the emergence of a prepeak---a signature of intermediate range order---that grows with increasing friction. 
We correlate the emergence of this peak to real space fluctuations in the local particle coordination number, which exhibits a grainy fluctuating field throughout the packing process that is retained in the final, mechanically stable state.
While the formation of the prepeak shows varying degrees of robustness to packing protocol changes, our results suggest that preparation history may be used to construct packings with variable large length scale structural properties. 

\end{abstract}

\maketitle

Mechanically stable packings of monodisperse spheres present as a model for studies of disordered, jammed systems. 
The macroscopic structure of frictionless jammed packings resembles those of other amorphous systems, such as glasses, inasmuch as there are traditional signatures of local structure~\cite{ohern2003}.
This is evinced by a primary, nearest-neighbor peak readily seen in the static structure factor, $S(q)$, for wavenumbers $q$ corresponding to \Silbert{the average inter-particle separation, which is commensurate with the typical particle size for short-range contact forces characteristic of dry granular materials.} 
At larger wavenumbers, $S(q)$ undergoes decaying oscillations that asymptote to unity~\cite{ohern2003}. 
For regular thermal systems, the small wavenumber limit of $S(q)$ relates to the mechanical properties of the system, i.e., $S(q \rightarrow 0) = S_{0} = \rho k_BT \chi_T$, in the thermodynamic limit, where  $\rho$ is the density of the system and $\chi_T$ is the isothermal susceptibility \cite{hansenmcdonald}.  
However, it is now widely accepted that at longer wavelengths---larger length scales---jammed particle packings exhibit suppressed density fluctuations compared to traditional liquid state theory, where a region of hyperuniformity persists, such that $S(q) \sim q$ \cite{donev2005,silbert2009,ikeda2017,jiao2021}.  

An in-between scenario is also not uncommon in liquids and structural glasses, including colloidal systems. 
At wavenumbers between the thermodynamic limit and the primary nearest-neighbor peak, evidence of structural correlations can sometimes emerge via the appearance of a prepeak---\Silbert{also known as the first sharp diffraction peak}: a feature in $S(q)$ that lies in the range $0 < q\sigma < 2\pi$, where $\sigma$ represents the characteristic particle diameter.
\Silbert{A prepeak in this $q$ range signifies intermediate range order (IRO) that is suggestive of a repeating motif spanning length scales of several, if not many, particle diameters. 
It is worth emphasizing that IRO, corresponding to large length scale structural features, predominately arises in systems with competing interactions over length scales of both short- and long-range character, or via directional bonding.  
These include metallic glasses~\cite{DLPrice_1989,Sheng2006}, ionic or molten liquids~\cite{Parmar2024}, colloidal systems~\cite{Godfrin2013,liu2019-2}, and network-forming glasses~\cite{Salmon2013}. 
For example, the repeating motif responsible for IRO observed in silica is due to interconnected tetrahedral structures formed by the directional Si-O bonds.
IRO has also been invoked as a controlling factor of packing efficiency and the dynamics in dense colloidal systems~\cite{yuan2021,singh2023}, such as stabilizing structures that develop during flow~\cite{silbert1999}.}

While the general and superficial structural features between dense liquids and jammed packings share many similarities~\cite{ohern2003,kamien2007,silbert2009,ikeda2015}, unlike liquids, and the other systems listed above, static granular packings must satisfy not only global but also local conditions of mechanical stability merely through local contact interactions. 
This leads to constraints on both the mean particle coordination number and the local number of nearest neighbors with which individual particles directly interact.
Such features are not usually qualified by typical averaging measures such as $S(q)$. 
Thus, the requirement that both local and global properties must result in stability begs the question as to whether additional structural organization does indeed exist within the jammed phase despite the lack of any long-range particle interactions present. 
It is precisely in this region of $S(q)$ where this work is focused. 
Here, we present static structure factor analyses that highlight the structural signatures in this intermediate range of length scales, for granular packings of frictionless and frictional particles using unprecedented large scale simulations of up to $\mathcal{O}(10^{8})$ particles.

To address these questions, we performed three-dimensional discrete element method simulations of spherical particles in LAMMPS~\cite{thompson2022}.
Particles interacted only on contact via frictional, damped, repulsive Hookean springs with spring constant $k$~\cite{cundall1979,silbert2001}, with rolling and twisting resistances modeled using Luding's technique~\cite{luding2008,santos2020}.
Friction coefficients $\mu_s$, $\mu_r$, and $\mu_t$ set the sliding, rolling, and twisting resistances, respectively.
Initial particle configurations were generated by randomly placing $N$ particles with diameter $\sigma$~\cite{dispersity_key} into a periodic cubic box with average particle density $\phi_0 = N\pi\sigma^3/6V_0$, where $V_0$ is the initial volume ($\phi_0 = 0.01$ unless otherwise stated).
Particle overlaps in the starting configurations were relaxed by running constant volume dynamics with viscous damping until the kinetic energy vanished.
Mechanically stable jammed packings were created by applying a small hydrostatic pressure $P$ to contract the volume using the isobaric-isoenthalpic ensemble, with $P\sigma/k=10^{-5}$ to probe the hard-particle limit of jamming; other simulation parameters closely followed~\citet{santos2020}. 
\Monti{The simulation time step size was set to $\approx 0.03\sqrt{m/k}$, where $m$ is the particle mass.}
The simulations proceeded until the internal and applied stress tensors matched and the kinetic energy per particle was small, typically $\mathcal{O}(10^{-12}k\sigma^2)$ or less.
The final jammed simulation boxes were triclinic with small tilt values on the order of $\sigma$.
Calculation of $S(q)$ for triclinic boxes was accomplished following~\citet{monti2023}.

\begin{figure}
\centering
\includegraphics[width=0.47\textwidth]{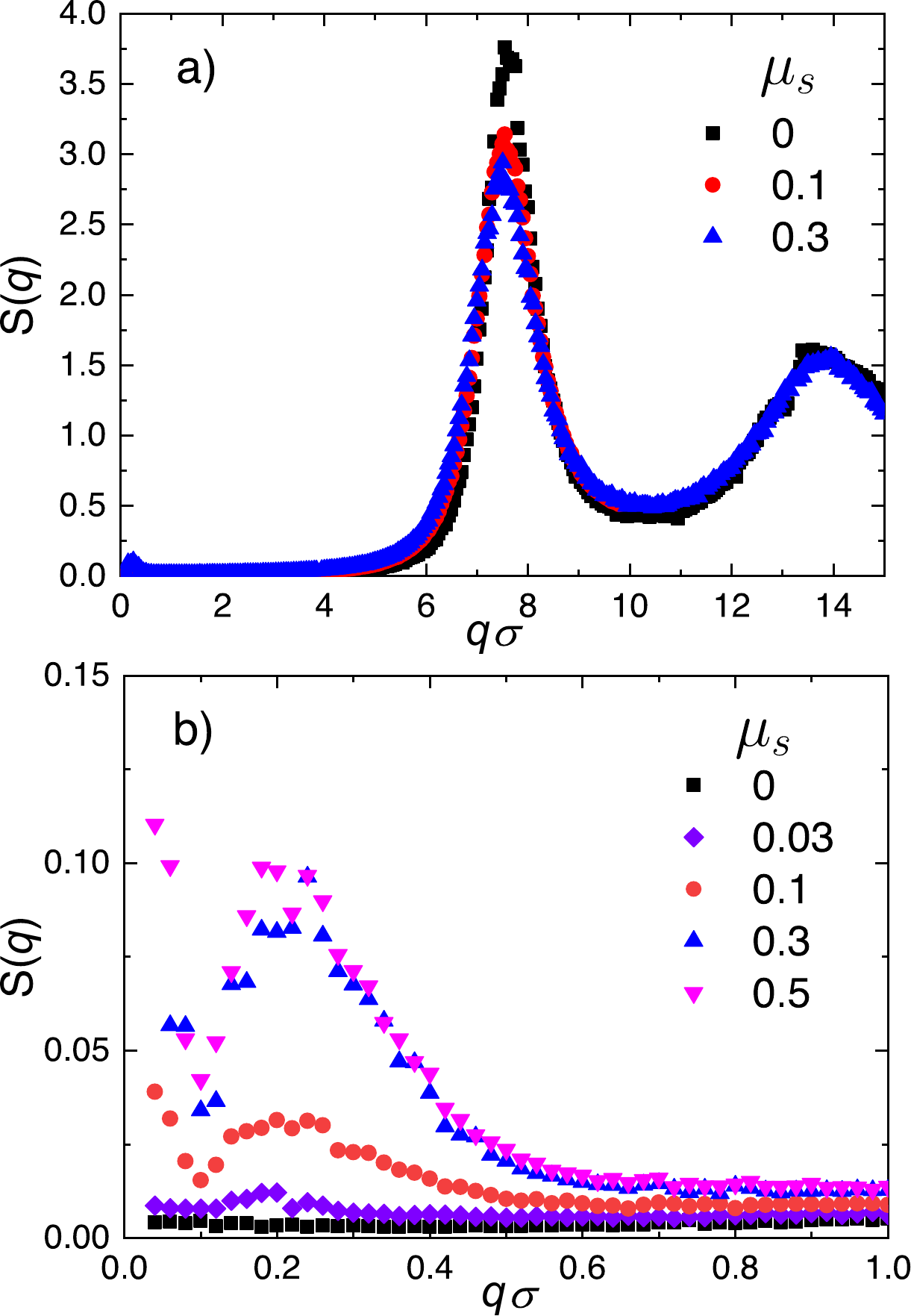}
\caption{(a) Static structure factor $S(q)$ for packings with $N\ = {10}^6$ particles at the indicated values of the sliding friction coefficient $\mu_s$. 
(b) $S(q)$ at low $q$ for packings with $N\ = {10}^7$ particles over a broader range of sliding friction coefficients $\mu_s$.
The associated $\phi$ values for increasing $\mu_s$ in b) are: 0.638, 0.624, 0.611, 0.595, and 0.590, respectively.}
\label{fig:fig1}
\end{figure}

Most features of the static structure factor for frictionless, monodisperse systems are recognizable, such as the positions and amplitudes of the primary and higher order peaks, and the $q\rightarrow\infty$ asymptotic behavior~\cite{ohern2003,donev2005,silbert2009}.
Figure~\ref{fig:fig1}a shows $S(q)$ for frictionless and frictional particle packings with $N = 10^6$ particles; the salient effect of friction is to diminish the primary peak, which is consistent with decreasing packing fraction and average particle coordination number as friction increases \cite{silbert2002-2}.
The corresponding volume fractions range from $\phi = 0.639$ for $\mu_s=0$ to $\phi = 0.594$ for $\mu_s = 0.3$.
However, for the large packings of frictional particles studied herein, there are clear signatures of structure on intermediate scales, i.e., on scales much larger than $\sigma$ and much smaller than the characteristic simulation length $\sim V^{1/3}$.
Figure~\ref{fig:fig1}b magnifies the low $q$ region for varying $\mu_s$ and reveals the emergence of a prepeak spanning a range of $q$ loosely corresponding to length scales of $10\sigma-60\sigma$, at its broadest.
$S(q)$ for frictionless, athermal packings has universally been found to increase monotonically with $q$ in this regime~\cite{ohern2003,donev2005,ikeda2015,ikeda2017,maher2023}; conversely, we highlight here that the prepeak for even a modest amount of sliding friction is distinguishable from frictionless data at similar $q$, despite only small changes to global packing metrics such as $\phi$ or the mean coordination number with nonzero friction~\cite{santos2020}.
The amplitude of the prepeak grows quickly with increasing $\mu_s$ before saturating for $\mu_s \geq 0.3$, beyond which additional sliding friction has limited impact~\cite{silbert2002-1}.

\begin{figure}
\centering
\includegraphics[width=0.47\textwidth]{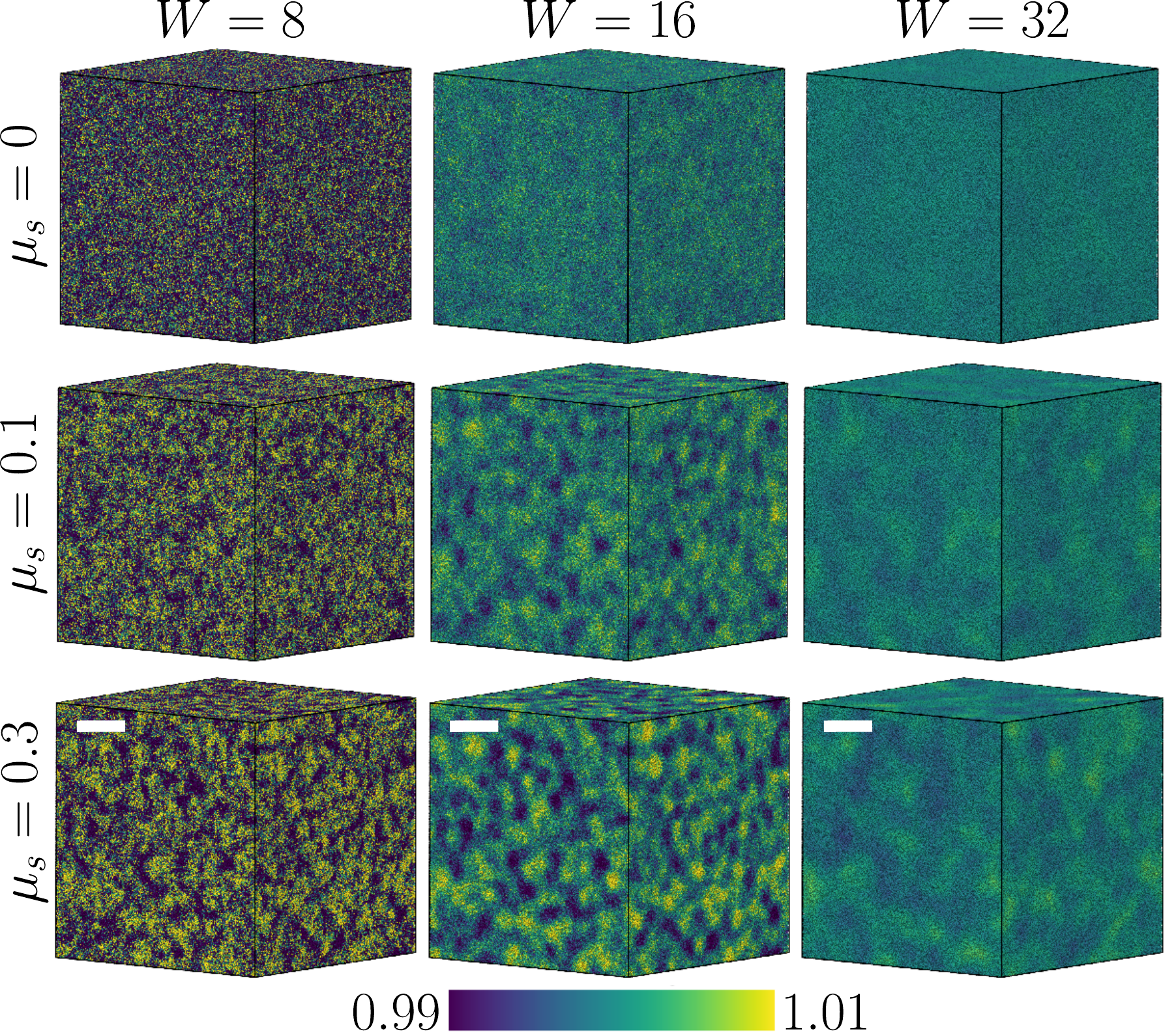}
\caption{Visualizations of the variation of particle coordination evaluated over a spherical window with diameter $W\sigma$, normalized by the coordination expected assuming uniform particle density, for three values of $\mu_{s}$. 
The color bar indicates coordination variations of $\pm1\%$. Each packing contains $N\ = 10^{7}$ particles, with the characteristic simulation length $\sim V^{1/3}\sim 200\sigma$. 
The superposed scale bar in the final row of images corresponds to $50\sigma$. 
The images were rendered in OVITO~\cite{ovito}.}
\label{fig:fig2}
\end{figure}

Structure over intermediate length scales leading to the observed IRO, can be demonstrated visually by quantifying variations of particle density.
Figure~\ref{fig:fig2} depicts fluctuations in the number of neighbors (coordination number) contained within a spherical window with diameter $W\sigma$ centered on each particle, normalized by the number of neighbors expected assuming the density within the window is $\phi W^3$~\cite{donev2005}.
This strategy picks out density variations on scales comparable to $W\sigma$, which manifest in $S(q)$ as peaks at $q\sigma\sim 2\pi/W$.
Considering $8 \leq W \leq 32$, the images indicate that frictionless packings exhibit no marked density fluctuations on scales much larger than $\sigma$, whereas the frictional packings show characteristic emergent structure, an effect that grows more prominent with increasing friction.
This range of window sizes corresponds to the upper half of wavenumbers where the prepeak is apparent in Fig.~\ref{fig:fig1}b, and is much smaller than the simulation length $V^{1/3}\sim 200\sigma$. 

\begin{figure}
\centering
\includegraphics[width=0.47\textwidth]{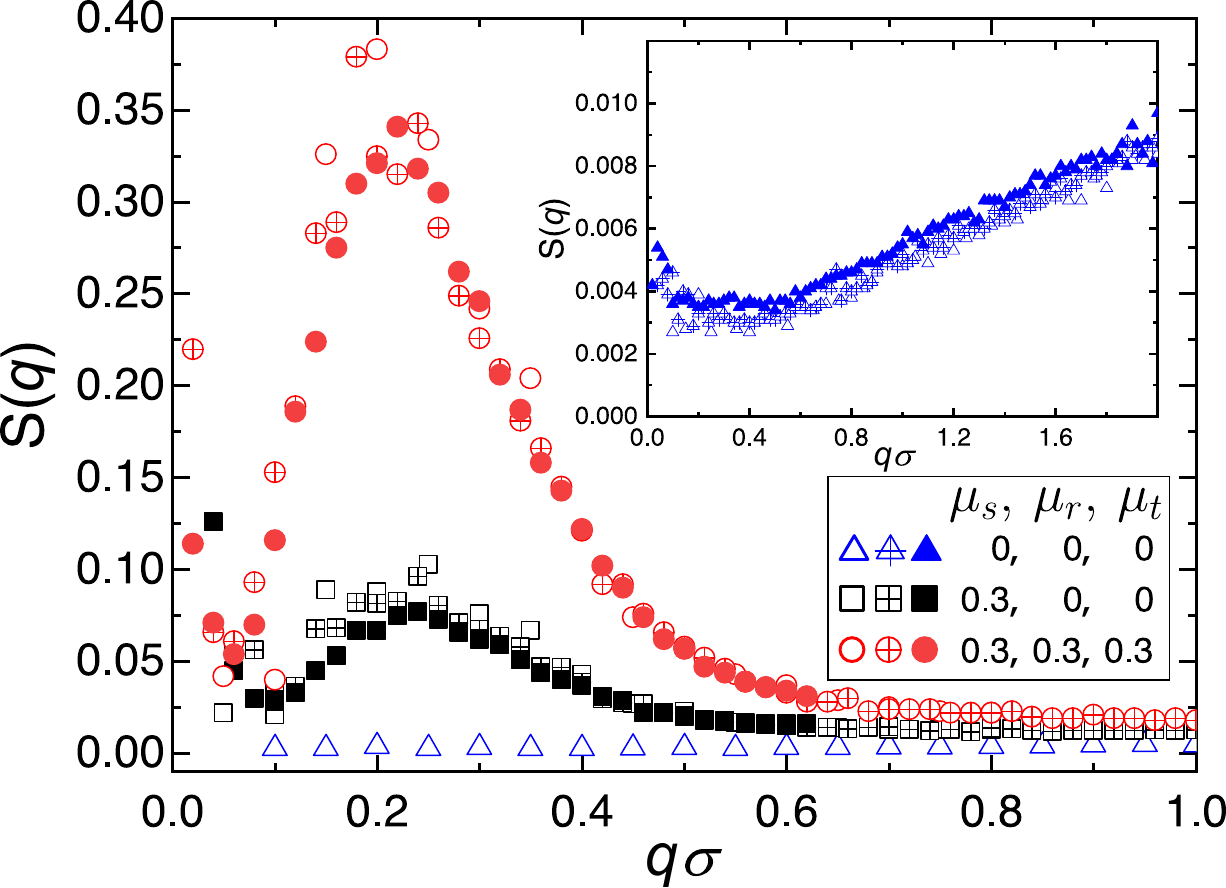}
\caption{$S(q)$ computed for packings with $N\ = 10^{6}$ (open), $10{^7}$ (hatched), and $8 \times 10{^7}$ (filled symbols) for varying $\mu_{s}$, $\mu_{r}$, and $\mu_{t}$ indicated in the legend. 
Inset: $S(q)$ for frictionless systems with symbols matching those in the main figure.}
\label{fig:fig3}
\end{figure}

The effect of introducing additional frictional mechanisms by including rolling and twisting resistances between particles is shown in Fig.~\ref{fig:fig3}.
For $\mu_s = \mu_r = \mu_t = 0.3$ ($\phi = 0.572$ for $N=10^7$), the prepeak spans a broader range of $q$ and more than triples in magnitude compared to the case with only sliding friction, though the peak maxima occur at nearly identical $q$.
\Silbert{Thus, addition of rolling and twisting friction modes results in a dramatic increase in the height of the prepeak while also producing significant prepeak broadening, suggesting that correlated particle domains become more varied in size---additional friction modes can support lower coordinated particles~\cite{santos2020}, which perhaps lends itself to a more varied range in stable structures---but because such domains become more abundant, this leads to enhanced correlations, resulting in the growth of the prepeak.}
Moreover, Fig.~\ref{fig:fig3} also demonstrates that the prepeak is independent of system size (provided the system is sufficiently large). Note that each increase in $N$ in the figure roughly halves the minimum accessible nonzero $q$ and drastically increases the number of wavenumbers comprising the prepeak.
Thence, the prepeak is not an artifact of the simulation interacting with its periodic images; however, we cannot rule out periodicity effects causing the upturn of $S(q)$ for the smallest $q\sigma\sim 2\pi\sigma/V^{1/3}$.

The inset of Fig.~\ref{fig:fig3} shows results for $S(q)$ for frictionless packings for systems up to size  $N = 8 \times 10{^7}$ (with $\phi = 0.638$). 
As found previously \cite{donev2005,silbert2009,ikeda2017}, $S(q)$ shows a linear relationship, i.e., $S(q) = A + Bq$, for intermediate wavenumbers. 
The intercept $A \approx 0.002$ is slightly larger than the earlier works reported, possibly owing to differing packing protocols. 
Additionally, a distinctive upturn in $S(q)$ appears at the smallest $q$ values for the larger systems, \Silbert{similar to measurements on amorphous silicon~\cite{steinhardt2013} and other studies of large packings~\cite{ikeda2017}, despite the systems here having been prepared at a packing pressure of $10^{-5}$, i.e., at the jamming transition.} 
Whether this upturn corresponds to the beginnings of a peak at very low $q$, similar to our findings for frictional particles, is an open question \Silbert{that would require study of even larger systems.}

\begin{figure}
\centering
\includegraphics[width=0.47\textwidth]{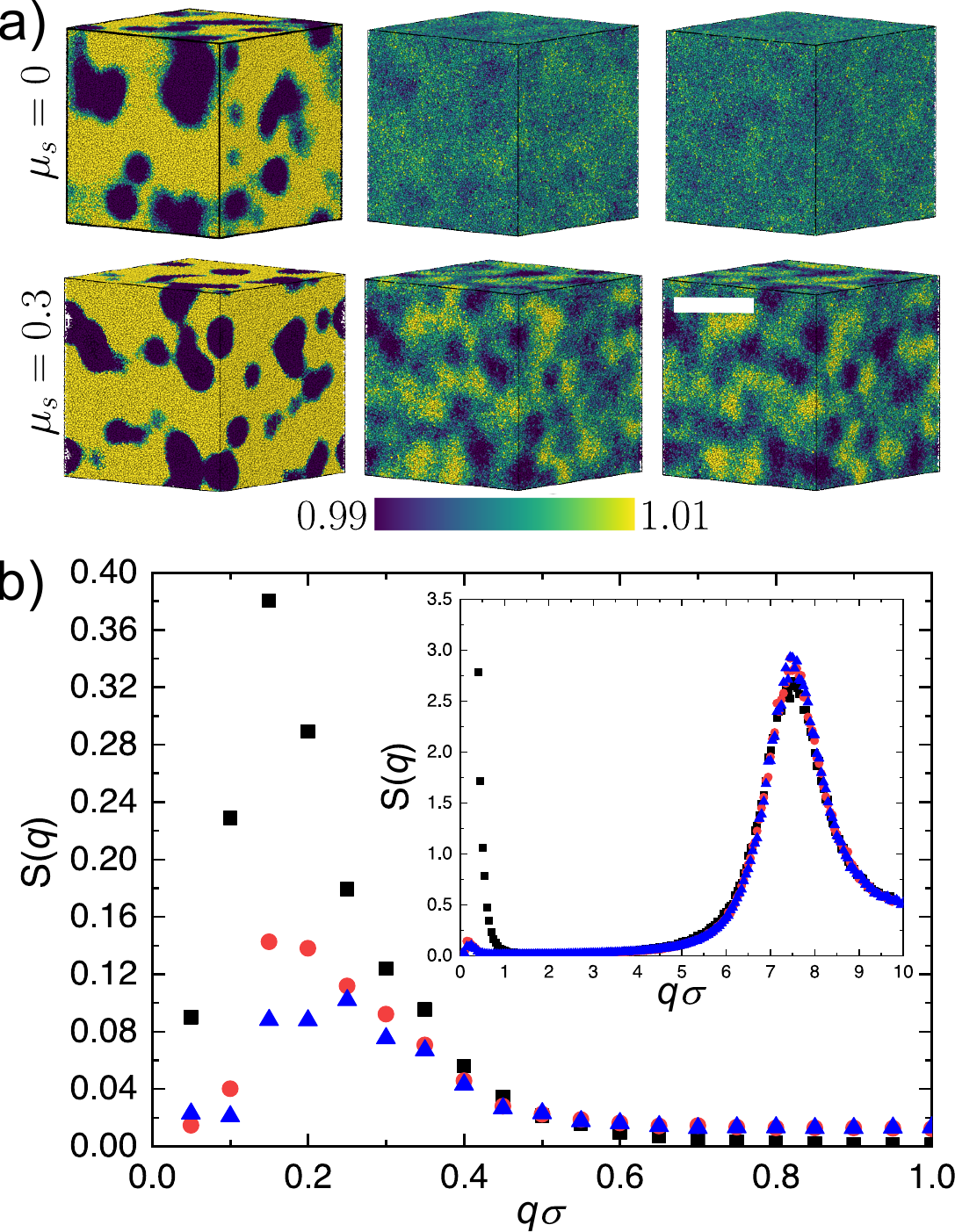}
\caption{(a) Visualizations of normalized coordination fluctuations evaluated with $W = 16$ for systems with $N\ =  10^{6}$ particles at the indicated values of $\mu_{s}$. 
The first two visualizations occur at $10^{5}$ and $\ 2\ \times 10^{5}$ time steps, and the final configurations at $6.6\times 10^{7}$ time steps for $\mu_{s} = 0$ and $8\times 10^6$ time steps for $\mu_{s} = 0.3$. 
The scale bar in the final image indicates $50\sigma$. 
(b) $S(q)$ at low $q$  for the $\mu_{s}=0.3$ configurations depicted in (a), shown in ascending order of simulation time as black squares, red circles, and blue triangles. 
The earliest simulation time data (black squares) are scaled by a factor of 0.02 to facilitate comparison of the low $q$ ranges (see inset). 
Inset: unscaled $S(q)$ over a larger range of $q$.}
\label{fig:fig4}
\end{figure}

\Silbert{Thus, it transpires that the formation of IRO is related to the formation of correlated fluctuating domains in particle coordination density. 
The development of these correlated domains can be examined through a time sequence of packing configurations.~\cite{vidoe_key}}
Figure~\ref{fig:fig4}a shows coordination variations for early, later, and final configurations for frictionless and frictional particles, both of which begin from the same dilute initial state.
The visualization scheme ($W = 16$) highlights large coordination variations that form over the first $10^5$ time steps of consolidation (from $\phi_0 = 0.01$ to $\phi \approx 0.600$ and $0.572$ for frictionless and frictional, respectively), but at $2\times 10^5$ time steps it is clear that the frictionless system anneals out the early-time structure, while the variations are more resilient in frictional systems.
Moreover, the second and third images appear similar for each case, implying that local particle rearrangements during the latter packing stages ($\phi \approx 0.629\rightarrow 0.639$ and $0.593\rightarrow0.594$) do not disrupt the IRO.

To quantify these observations, $S(q)$ for the frictional configurations of Fig.~\ref{fig:fig4}a are shown in Fig.~\ref{fig:fig4}b.
After $10^5$ time steps, the prepeak is present over roughly the same $q$ range as in the final configuration, albeit with a much larger amplitude; to facilitate comparison with later times, the prepeak at $10^5$ time steps is scaled by $0.02$ in the main figure, while the inset plots the unscaled data, demonstrating that the prepeak is initially comparable to the primary neighbor peak in magnitude.
Between $2\times 10^5$ time steps and the final configuration, there is a slight reduction and shift of the prepeak, implying that the structure is essentially locked in at early packing stages.

\begin{figure}
\centering
\includegraphics[width=0.47\textwidth]{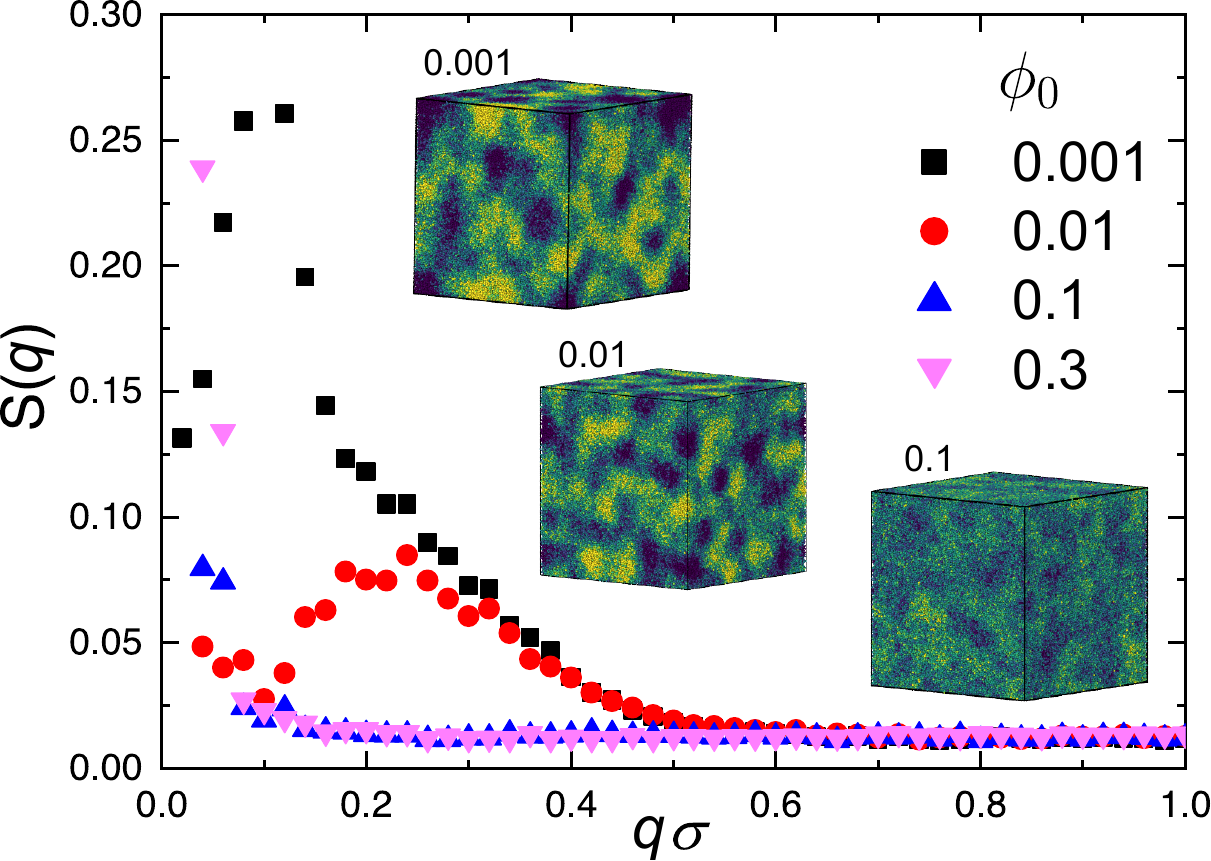}
\caption{$S(q)$ at low $q$ for packings with the indicated values of initial volume fraction $\phi_0$ for a system of $N=10^7$ particles with $\mu_s=0.3$. 
Snapshots of several of the corresponding normalized coordination fluctuations for $N =10^6$, evaluated with $W = 16$ are included. 
Particle coloration is identical to Fig.~\ref{fig:fig4}a.}
\label{fig:fig5}
\end{figure}

The IRO exhibited above is dependent on both initial configuration and packing protocol.
Focusing on the role of initial density, $\phi_0$, Fig.~\ref{fig:fig5} compares $S(q)$ for small $q$ and coordination variation images for a range of $\phi_0$.
Final packing volume fractions for $\phi_0 = 0.001$ and 0.1 are similar to $\phi_0 = 0.01$, with $\phi \approx 0.595$, whereas the $\phi_0 = 0.3$ system packs more loosely to $\phi = 0.590$.
Compared to $\phi_0 = 0.01$, the prepeak for $\phi_0 = 0.001$ is amplified and shifted to lower $q$ (larger length scale).
Packings starting with $\phi_0 = 0.1$ and 0.3 do not exhibit IRO, but they do exhibit strong peaks at the lowest nonzero $q$.
The dichotomy across $\phi_0$ helps explain the physical origin of the density fluctuations: starting from a dilute, randomized state, volumetric contraction produces both clumps of particles and depleted regions.
Higher density initial configurations do not contract as far and so are less clumpy at the outset, and have stronger correlations between particle positions arising from the overlap-removing procedure during initialization.
Indeed, the radial distribution functions for the initial configurations (not shown) reveal no pronounced liquid-like ordering for $\phi_0 = 0.001$ and $0.01$, but a clear contact peak forms for $\phi_0 = 0.1$ and additional secondary structure appears for $\phi_0 = 0.3$.
Among other factors, including kinetic energy introduced during packing, the degree of homogeneity at early stages may play an important role in producing IRO.


Protocol dependence in frictional, non-cohesive granular materials is well known~\cite{nowak1997,santomaso2003,landry2003,somfai2007,song2008,silbert2010,bililign2019,luding2016,santos2024}.
We found that the IRO described in this work could be reduced or eliminated through several means, including: adding a strong damping contribution to the barostat (thereby modifying the Nos\'e-Hoover formalism); over-compressing the packing at higher $P$ before reducing to the nominal $P$; or, employing the Berendsen barostat~\cite{berendsen1984} as an alternative.
Raising $P$ by one or two orders of magnitude while maintaining our original packing strategy did not entirely remove the prepeak. 
It is also conceivable that volume-controlled packing strategies will suppress the fragile structures that produce the prepeak, considering the challenges they pose for jamming particles at low pressures~\cite{silbert2010,santos2024}, although we did not test this supposition here.
\Monti{Finally, the structure apparent in strongly frictional packings (e.g., $N = 10^7$, $\mu_s = \mu_r = \mu_t = 0.3$) was considerably weakened by subsequently applying a unit strain of low-rate simple shear---shear dilation reduced $\phi$ by $\sim 0.01$ in this instance~\cite{srivastava2021b}.}

\Monti{As a counterpoint to the isotropic compression protocol without gravity described above, packings can instead be constructed by pouring particles under gravity onto a supporting surface.
The End Matter describes results of a limited number of pouring simulations that produced packings lacking a notable $S(q)$ prepeak, but with coordination variations exhibiting similar trends with increasing friction as those depicted in Fig.~\ref{fig:fig2}.
The pouring approach provides an additional degree of realism to granular packing simulations and warrants greater consideration in future studies.}


\Silbert{To summarize, static structure factor calculations for large jammed packings of frictional monodisperse spheres revealed the presence of intermediate range order through the identification of a prepeak, or first diffraction peak, at wavenumbers below that of the nearest-neighbor primary peak. 
This feature is unexpected for systems that are governed purely by contact forces, given that the presence of large length scale correlations typically appear only in systems with longer-range interactions.} 
These correlations correspond to intermediate length scales of order 10 to 60 times the particle diameter, which appear to emerge from particle density fluctuations at early packing stages. 
This intermediate structure is absent for packings of frictionless particles, which instead displayed linear behavior over a limited range of wavenumbers with an upturn at the lowest $q$, consistent with other recent experiments and simulations. 
However, it does seem apparent that packing protocol and initial conditions are crucial for preserving intermediate range order in jammed configurations, and the level of correlations in the connectivity of the packing aids in tuning what heterogeneous structures are achievable.


I.S. acknowledges support from the U.S. Department of Energy (DOE), Office of Science, Office of Advanced Scientific Computing Research, Applied Mathematics Program under Contract No. DE-AC02-05CH11231.
This work was performed in part at the Center for Integrated Nanotechnologies, a U.S. DOE and Office of Basic Energy Sciences user facility.
Sandia National Laboratories is a multimission laboratory managed and operated by National Technology \& Engineering Solutions of Sandia, LLC, a wholly owned subsidiary of Honeywell International Inc., for the U.S. DOE’s National Nuclear Security Administration under contract DE-NA0003525.
This paper describes objective technical results and analysis. 
Any subjective views or opinions that might be expressed in the paper do not necessarily represent the views of the U.S. DOE or the U.S. Government.

\bibliography{bibfile}

\begin{thebibliography}{41}%
\makeatletter
\providecommand \@ifxundefined [1]{%
 \@ifx{#1\undefined}
}%
\providecommand \@ifnum [1]{%
 \ifnum #1\expandafter \@firstoftwo
 \else \expandafter \@secondoftwo
 \fi
}%
\providecommand \@ifx [1]{%
 \ifx #1\expandafter \@firstoftwo
 \else \expandafter \@secondoftwo
 \fi
}%
\providecommand \natexlab [1]{#1}%
\providecommand \enquote  [1]{``#1''}%
\providecommand \bibnamefont  [1]{#1}%
\providecommand \bibfnamefont [1]{#1}%
\providecommand \citenamefont [1]{#1}%
\providecommand \href@noop [0]{\@secondoftwo}%
\providecommand \href [0]{\begingroup \@sanitize@url \@href}%
\providecommand \@href[1]{\@@startlink{#1}\@@href}%
\providecommand \@@href[1]{\endgroup#1\@@endlink}%
\providecommand \@sanitize@url [0]{\catcode `\\12\catcode `\$12\catcode
  `\&12\catcode `\#12\catcode `\^12\catcode `\_12\catcode `\%12\relax}%
\providecommand \@@startlink[1]{}%
\providecommand \@@endlink[0]{}%
\providecommand \url  [0]{\begingroup\@sanitize@url \@url }%
\providecommand \@url [1]{\endgroup\@href {#1}{\urlprefix }}%
\providecommand \urlprefix  [0]{URL }%
\providecommand \Eprint [0]{\href }%
\providecommand \doibase [0]{https://doi.org/}%
\providecommand \selectlanguage [0]{\@gobble}%
\providecommand \bibinfo  [0]{\@secondoftwo}%
\providecommand \bibfield  [0]{\@secondoftwo}%
\providecommand \translation [1]{[#1]}%
\providecommand \BibitemOpen [0]{}%
\providecommand \bibitemStop [0]{}%
\providecommand \bibitemNoStop [0]{.\EOS\space}%
\providecommand \EOS [0]{\spacefactor3000\relax}%
\providecommand \BibitemShut  [1]{\csname bibitem#1\endcsname}%
\let\auto@bib@innerbib\@empty
\bibitem [{\citenamefont {O'Hern}\ \emph {et~al.}(2003)\citenamefont {O'Hern},
  \citenamefont {Silbert}, \citenamefont {Liu},\ and\ \citenamefont
  {Nagel}}]{ohern2003}%
  \BibitemOpen
  \bibfield  {author} {\bibinfo {author} {\bibfnamefont {C.~S.}\ \bibnamefont
  {O'Hern}}, \bibinfo {author} {\bibfnamefont {L.~E.}\ \bibnamefont {Silbert}},
  \bibinfo {author} {\bibfnamefont {A.~J.}\ \bibnamefont {Liu}},\ and\ \bibinfo
  {author} {\bibfnamefont {S.~R.}\ \bibnamefont {Nagel}},\ }\href
  {https://doi.org/10.1103/PhysRevE.68.011306} {\bibfield  {journal} {\bibinfo
  {journal} {Phys. Rev. E}\ }\textbf {\bibinfo {volume} {68}},\ \bibinfo
  {pages} {011306} (\bibinfo {year} {2003})}\BibitemShut {NoStop}%
\bibitem [{\citenamefont {Hansen}\ and\ \citenamefont
  {McDonald}(2013)}]{hansenmcdonald}%
  \BibitemOpen
  \bibfield  {author} {\bibinfo {author} {\bibfnamefont {J.-P.}\ \bibnamefont
  {Hansen}}\ and\ \bibinfo {author} {\bibfnamefont {I.~R.}\ \bibnamefont
  {McDonald}},\ }\href
  {https://doi.org/https://doi.org/10.1016/C2010-0-66723-X} {\emph {\bibinfo
  {title} {Theory of Simple Liquids}}},\ \bibinfo {edition} {4th}\ ed.\
  (\bibinfo  {publisher} {Academic Press},\ \bibinfo {address} {Oxford},\
  \bibinfo {year} {2013})\BibitemShut {NoStop}%
\bibitem [{\citenamefont {Donev}\ \emph {et~al.}(2005)\citenamefont {Donev},
  \citenamefont {Stillinger},\ and\ \citenamefont {Torquato}}]{donev2005}%
  \BibitemOpen
  \bibfield  {author} {\bibinfo {author} {\bibfnamefont {A.}~\bibnamefont
  {Donev}}, \bibinfo {author} {\bibfnamefont {F.~H.}\ \bibnamefont
  {Stillinger}},\ and\ \bibinfo {author} {\bibfnamefont {S.}~\bibnamefont
  {Torquato}},\ }\href {https://doi.org/10.1103/PhysRevLett.95.090604}
  {\bibfield  {journal} {\bibinfo  {journal} {Phys. Rev. Lett.}\ }\textbf
  {\bibinfo {volume} {95}},\ \bibinfo {pages} {090604} (\bibinfo {year}
  {2005})}\BibitemShut {NoStop}%
\bibitem [{\citenamefont {Silbert}\ and\ \citenamefont
  {Silbert}(2009)}]{silbert2009}%
  \BibitemOpen
  \bibfield  {author} {\bibinfo {author} {\bibfnamefont {L.~E.}\ \bibnamefont
  {Silbert}}\ and\ \bibinfo {author} {\bibfnamefont {M.}~\bibnamefont
  {Silbert}},\ }\href {https://doi.org/10.1103/PhysRevE.80.041304} {\bibfield
  {journal} {\bibinfo  {journal} {Phys. Rev. E}\ }\textbf {\bibinfo {volume}
  {80}},\ \bibinfo {pages} {041304} (\bibinfo {year} {2009})}\BibitemShut
  {NoStop}%
\bibitem [{\citenamefont {Ikeda}\ \emph {et~al.}(2017)\citenamefont {Ikeda},
  \citenamefont {Berthier},\ and\ \citenamefont {Parisi}}]{ikeda2017}%
  \BibitemOpen
  \bibfield  {author} {\bibinfo {author} {\bibfnamefont {A.}~\bibnamefont
  {Ikeda}}, \bibinfo {author} {\bibfnamefont {L.}~\bibnamefont {Berthier}},\
  and\ \bibinfo {author} {\bibfnamefont {G.}~\bibnamefont {Parisi}},\ }\href
  {https://doi.org/10.1103/PhysRevE.95.052125} {\bibfield  {journal} {\bibinfo
  {journal} {Phys. Rev. E}\ }\textbf {\bibinfo {volume} {95}},\ \bibinfo
  {pages} {052125} (\bibinfo {year} {2017})}\BibitemShut {NoStop}%
\bibitem [{\citenamefont {Yuan}\ \emph
  {et~al.}(2021{\natexlab{a}})\citenamefont {Yuan}, \citenamefont {Jiao},
  \citenamefont {Wang},\ and\ \citenamefont {Li}}]{jiao2021}%
  \BibitemOpen
  \bibfield  {author} {\bibinfo {author} {\bibfnamefont {Y.}~\bibnamefont
  {Yuan}}, \bibinfo {author} {\bibfnamefont {Y.}~\bibnamefont {Jiao}}, \bibinfo
  {author} {\bibfnamefont {Y.}~\bibnamefont {Wang}},\ and\ \bibinfo {author}
  {\bibfnamefont {S.}~\bibnamefont {Li}},\ }\href
  {https://doi.org/10.1103/PhysRevResearch.3.033084} {\bibfield  {journal}
  {\bibinfo  {journal} {Phys. Rev. Res.}\ }\textbf {\bibinfo {volume} {3}},\
  \bibinfo {pages} {033084} (\bibinfo {year} {2021}{\natexlab{a}})}\BibitemShut
  {NoStop}%
\bibitem [{\citenamefont {Price}\ \emph {et~al.}(1989)\citenamefont {Price},
  \citenamefont {Moss}, \citenamefont {Reijers}, \citenamefont {Saboungi},\
  and\ \citenamefont {Susman}}]{DLPrice_1989}%
  \BibitemOpen
  \bibfield  {author} {\bibinfo {author} {\bibfnamefont {D.~L.}\ \bibnamefont
  {Price}}, \bibinfo {author} {\bibfnamefont {S.~C.}\ \bibnamefont {Moss}},
  \bibinfo {author} {\bibfnamefont {R.}~\bibnamefont {Reijers}}, \bibinfo
  {author} {\bibfnamefont {M.~L.}\ \bibnamefont {Saboungi}},\ and\ \bibinfo
  {author} {\bibfnamefont {S.}~\bibnamefont {Susman}},\ }\href
  {https://doi.org/10.1088/0953-8984/1/5/017} {\bibfield  {journal} {\bibinfo
  {journal} {J. Phys.: Cond. Matt.}\ }\textbf {\bibinfo {volume} {1}},\
  \bibinfo {pages} {1005} (\bibinfo {year} {1989})}\BibitemShut {NoStop}%
\bibitem [{\citenamefont {Sheng}\ \emph {et~al.}(2006)\citenamefont {Sheng},
  \citenamefont {Luo}, \citenamefont {Alamgir}, \citenamefont {Bai},\ and\
  \citenamefont {Ma}}]{Sheng2006}%
  \BibitemOpen
  \bibfield  {author} {\bibinfo {author} {\bibfnamefont {H.~W.}\ \bibnamefont
  {Sheng}}, \bibinfo {author} {\bibfnamefont {W.~K.}\ \bibnamefont {Luo}},
  \bibinfo {author} {\bibfnamefont {F.~M.}\ \bibnamefont {Alamgir}}, \bibinfo
  {author} {\bibfnamefont {J.~M.}\ \bibnamefont {Bai}},\ and\ \bibinfo {author}
  {\bibfnamefont {E.}~\bibnamefont {Ma}},\ }\href
  {https://doi.org/10.1038/nature04421} {\bibfield  {journal} {\bibinfo
  {journal} {Nature}\ }\textbf {\bibinfo {volume} {439}},\ \bibinfo {pages}
  {419} (\bibinfo {year} {2006})}\BibitemShut {NoStop}%
\bibitem [{\citenamefont {Parmar}\ \emph {et~al.}(2024)\citenamefont {Parmar},
  \citenamefont {Dean}, \citenamefont {Do}, \citenamefont {Browning},
  \citenamefont {Klein}, \citenamefont {Gurkan},\ and\ \citenamefont
  {McDaniel}}]{Parmar2024}%
  \BibitemOpen
  \bibfield  {author} {\bibinfo {author} {\bibfnamefont {S.~M.}\ \bibnamefont
  {Parmar}}, \bibinfo {author} {\bibfnamefont {W.}~\bibnamefont {Dean}},
  \bibinfo {author} {\bibfnamefont {C.}~\bibnamefont {Do}}, \bibinfo {author}
  {\bibfnamefont {J.~F.}\ \bibnamefont {Browning}}, \bibinfo {author}
  {\bibfnamefont {J.~M.}\ \bibnamefont {Klein}}, \bibinfo {author}
  {\bibfnamefont {B.~E.}\ \bibnamefont {Gurkan}},\ and\ \bibinfo {author}
  {\bibfnamefont {J.~G.}\ \bibnamefont {McDaniel}},\ }\href
  {https://doi.org/doi: 10.1021/acs.jpcb.4c06255.} {\bibfield  {journal}
  {\bibinfo  {journal} {J Phys Chem B.}\ }\textbf {\bibinfo {volume} {128}},\
  \bibinfo {pages} {11313} (\bibinfo {year} {2024})}\BibitemShut {NoStop}%
\bibitem [{\citenamefont {Godfrin}\ \emph {et~al.}(2013)\citenamefont
  {Godfrin}, \citenamefont {Castañeda-Priego}, \citenamefont {Liu},\ and\
  \citenamefont {Wagner}}]{Godfrin2013}%
  \BibitemOpen
  \bibfield  {author} {\bibinfo {author} {\bibfnamefont {P.~D.}\ \bibnamefont
  {Godfrin}}, \bibinfo {author} {\bibfnamefont {R.}~\bibnamefont
  {Castañeda-Priego}}, \bibinfo {author} {\bibfnamefont {Y.}~\bibnamefont
  {Liu}},\ and\ \bibinfo {author} {\bibfnamefont {N.~J.}\ \bibnamefont
  {Wagner}},\ }\href {https://doi.org/10.1063/1.4824487} {\bibfield  {journal}
  {\bibinfo  {journal} {J. Chem. Phys.}\ }\textbf {\bibinfo {volume} {139}},\
  \bibinfo {pages} {154904} (\bibinfo {year} {2013})}\BibitemShut {NoStop}%
\bibitem [{\citenamefont {Liu}\ and\ \citenamefont {Xi}(2019)}]{liu2019-2}%
  \BibitemOpen
  \bibfield  {author} {\bibinfo {author} {\bibfnamefont {Y.}~\bibnamefont
  {Liu}}\ and\ \bibinfo {author} {\bibfnamefont {Y.}~\bibnamefont {Xi}},\
  }\href {https://doi.org/https://doi.org/10.1016/j.cocis.2019.01.016}
  {\bibfield  {journal} {\bibinfo  {journal} {Current Opinion in Colloid \&
  Interface Science}\ }\textbf {\bibinfo {volume} {39}},\ \bibinfo {pages}
  {123} (\bibinfo {year} {2019})}\BibitemShut {NoStop}%
\bibitem [{\citenamefont {Salmon}\ and\ \citenamefont
  {Zeidler}(2013)}]{Salmon2013}%
  \BibitemOpen
  \bibfield  {author} {\bibinfo {author} {\bibfnamefont {P.~S.}\ \bibnamefont
  {Salmon}}\ and\ \bibinfo {author} {\bibfnamefont {A.}~\bibnamefont
  {Zeidler}},\ }\href {https://doi.org/10.1039/C3CP51741A} {\bibfield
  {journal} {\bibinfo  {journal} {Phys. Chem. Chem. Phys.}\ }\textbf {\bibinfo
  {volume} {15}},\ \bibinfo {pages} {15286} (\bibinfo {year}
  {2013})}\BibitemShut {NoStop}%
\bibitem [{\citenamefont {Yuan}\ \emph
  {et~al.}(2021{\natexlab{b}})\citenamefont {Yuan}, \citenamefont {Zhang},
  \citenamefont {Kob},\ and\ \citenamefont {Wang}}]{yuan2021}%
  \BibitemOpen
  \bibfield  {author} {\bibinfo {author} {\bibfnamefont {H.}~\bibnamefont
  {Yuan}}, \bibinfo {author} {\bibfnamefont {Z.}~\bibnamefont {Zhang}},
  \bibinfo {author} {\bibfnamefont {W.}~\bibnamefont {Kob}},\ and\ \bibinfo
  {author} {\bibfnamefont {Y.}~\bibnamefont {Wang}},\ }\href
  {https://doi.org/10.1103/PhysRevLett.127.278001} {\bibfield  {journal}
  {\bibinfo  {journal} {Phys. Rev. Lett.}\ }\textbf {\bibinfo {volume} {127}},\
  \bibinfo {pages} {278001} (\bibinfo {year} {2021}{\natexlab{b}})}\BibitemShut
  {NoStop}%
\bibitem [{\citenamefont {Singh}\ \emph {et~al.}(2023)\citenamefont {Singh},
  \citenamefont {Zhang}, \citenamefont {Sood}, \citenamefont {Kob},\ and\
  \citenamefont {Ganapathy}}]{singh2023}%
  \BibitemOpen
  \bibfield  {author} {\bibinfo {author} {\bibfnamefont {N.}~\bibnamefont
  {Singh}}, \bibinfo {author} {\bibfnamefont {Z.}~\bibnamefont {Zhang}},
  \bibinfo {author} {\bibfnamefont {A.~K.}\ \bibnamefont {Sood}}, \bibinfo
  {author} {\bibfnamefont {W.}~\bibnamefont {Kob}},\ and\ \bibinfo {author}
  {\bibfnamefont {R.}~\bibnamefont {Ganapathy}},\ }\href
  {https://doi.org/10.1073/pnas.2300923120} {\bibfield  {journal} {\bibinfo
  {journal} {PNAS}\ }\textbf {\bibinfo {volume} {120}},\ \bibinfo {pages}
  {e2300923120} (\bibinfo {year} {2023})}\BibitemShut {NoStop}%
\bibitem [{\citenamefont {Silbert}\ \emph {et~al.}(1999)\citenamefont
  {Silbert}, \citenamefont {Melrose},\ and\ \citenamefont
  {Ball}}]{silbert1999}%
  \BibitemOpen
  \bibfield  {author} {\bibinfo {author} {\bibfnamefont {L.~E.}\ \bibnamefont
  {Silbert}}, \bibinfo {author} {\bibfnamefont {J.~R.}\ \bibnamefont
  {Melrose}},\ and\ \bibinfo {author} {\bibfnamefont {R.~C.}\ \bibnamefont
  {Ball}},\ }\href {https://doi.org/10.1080/00268979909483110} {\bibfield
  {journal} {\bibinfo  {journal} {Molecular Physics}\ }\textbf {\bibinfo
  {volume} {96}},\ \bibinfo {pages} {1667} (\bibinfo {year}
  {1999})}\BibitemShut {NoStop}%
\bibitem [{\citenamefont {Kamien}\ and\ \citenamefont
  {Liu}(2007)}]{kamien2007}%
  \BibitemOpen
  \bibfield  {author} {\bibinfo {author} {\bibfnamefont {R.~D.}\ \bibnamefont
  {Kamien}}\ and\ \bibinfo {author} {\bibfnamefont {A.~J.}\ \bibnamefont
  {Liu}},\ }\href {https://doi.org/10.1103/PhysRevLett.99.155501} {\bibfield
  {journal} {\bibinfo  {journal} {Phys. Rev. Lett.}\ }\textbf {\bibinfo
  {volume} {99}},\ \bibinfo {pages} {155501} (\bibinfo {year}
  {2007})}\BibitemShut {NoStop}%
\bibitem [{\citenamefont {Ikeda}\ and\ \citenamefont
  {Berthier}(2015)}]{ikeda2015}%
  \BibitemOpen
  \bibfield  {author} {\bibinfo {author} {\bibfnamefont {A.}~\bibnamefont
  {Ikeda}}\ and\ \bibinfo {author} {\bibfnamefont {L.}~\bibnamefont
  {Berthier}},\ }\href {https://doi.org/10.1103/PhysRevE.92.012309} {\bibfield
  {journal} {\bibinfo  {journal} {Phys. Rev. E}\ }\textbf {\bibinfo {volume}
  {92}},\ \bibinfo {pages} {012309} (\bibinfo {year} {2015})}\BibitemShut
  {NoStop}%
\bibitem [{\citenamefont {Thompson}\ \emph {et~al.}(2022)\citenamefont
  {Thompson}, \citenamefont {Aktulga}, \citenamefont {Berger}, \citenamefont
  {Bolintineanu}, \citenamefont {Brown}, \citenamefont {Crozier}, \citenamefont
  {{in 't Veld}}, \citenamefont {Kohlmeyer}, \citenamefont {Moore},
  \citenamefont {Nguyen}, \citenamefont {Shan}, \citenamefont {Stevens},
  \citenamefont {Tranchida}, \citenamefont {Trott},\ and\ \citenamefont
  {Plimpton}}]{thompson2022}%
  \BibitemOpen
  \bibfield  {author} {\bibinfo {author} {\bibfnamefont {A.~P.}\ \bibnamefont
  {Thompson}}, \bibinfo {author} {\bibfnamefont {H.~M.}\ \bibnamefont
  {Aktulga}}, \bibinfo {author} {\bibfnamefont {R.}~\bibnamefont {Berger}},
  \bibinfo {author} {\bibfnamefont {D.~S.}\ \bibnamefont {Bolintineanu}},
  \bibinfo {author} {\bibfnamefont {W.~M.}\ \bibnamefont {Brown}}, \bibinfo
  {author} {\bibfnamefont {P.~S.}\ \bibnamefont {Crozier}}, \bibinfo {author}
  {\bibfnamefont {P.~J.}\ \bibnamefont {{in 't Veld}}}, \bibinfo {author}
  {\bibfnamefont {A.}~\bibnamefont {Kohlmeyer}}, \bibinfo {author}
  {\bibfnamefont {S.~G.}\ \bibnamefont {Moore}}, \bibinfo {author}
  {\bibfnamefont {T.~D.}\ \bibnamefont {Nguyen}}, \bibinfo {author}
  {\bibfnamefont {R.}~\bibnamefont {Shan}}, \bibinfo {author} {\bibfnamefont
  {M.~J.}\ \bibnamefont {Stevens}}, \bibinfo {author} {\bibfnamefont
  {J.}~\bibnamefont {Tranchida}}, \bibinfo {author} {\bibfnamefont
  {C.}~\bibnamefont {Trott}},\ and\ \bibinfo {author} {\bibfnamefont {S.~J.}\
  \bibnamefont {Plimpton}},\ }\href {https://doi.org/10.1016/j.cpc.2021.108171}
  {\bibfield  {journal} {\bibinfo  {journal} {Comput. Phys. Comm.}\ }\textbf
  {\bibinfo {volume} {271}},\ \bibinfo {pages} {108171} (\bibinfo {year}
  {2022})}\BibitemShut {NoStop}%
\bibitem [{\citenamefont {Cundall}\ and\ \citenamefont
  {Strack}(1979)}]{cundall1979}%
  \BibitemOpen
  \bibfield  {author} {\bibinfo {author} {\bibfnamefont {P.~A.}\ \bibnamefont
  {Cundall}}\ and\ \bibinfo {author} {\bibfnamefont {O.~D.~L.}\ \bibnamefont
  {Strack}},\ }\href {https://doi.org/10.1680/geot.1979.29.1.47} {\bibfield
  {journal} {\bibinfo  {journal} {G\'eotechnique}\ }\textbf {\bibinfo {volume}
  {29}},\ \bibinfo {pages} {47} (\bibinfo {year} {1979})}\BibitemShut {NoStop}%
\bibitem [{\citenamefont {Silbert}\ \emph {et~al.}(2001)\citenamefont
  {Silbert}, \citenamefont {Erta\c{s}}, \citenamefont {Grest}, \citenamefont
  {Halsey}, \citenamefont {Levine},\ and\ \citenamefont
  {Plimpton}}]{silbert2001}%
  \BibitemOpen
  \bibfield  {author} {\bibinfo {author} {\bibfnamefont {L.~E.}\ \bibnamefont
  {Silbert}}, \bibinfo {author} {\bibfnamefont {D.}~\bibnamefont {Erta\c{s}}},
  \bibinfo {author} {\bibfnamefont {G.~S.}\ \bibnamefont {Grest}}, \bibinfo
  {author} {\bibfnamefont {T.~C.}\ \bibnamefont {Halsey}}, \bibinfo {author}
  {\bibfnamefont {D.}~\bibnamefont {Levine}},\ and\ \bibinfo {author}
  {\bibfnamefont {S.~J.}\ \bibnamefont {Plimpton}},\ }\href
  {https://doi.org/10.1103/PhysRevE.64.051302} {\bibfield  {journal} {\bibinfo
  {journal} {Phys. Rev. E}\ }\textbf {\bibinfo {volume} {64}},\ \bibinfo
  {pages} {051302} (\bibinfo {year} {2001})}\BibitemShut {NoStop}%
\bibitem [{\citenamefont {Luding}(2008)}]{luding2008}%
  \BibitemOpen
  \bibfield  {author} {\bibinfo {author} {\bibfnamefont {S.}~\bibnamefont
  {Luding}},\ }\href
  {https://doi.org/https://doi.org/10.1007/s10035-008-0099-x} {\bibfield
  {journal} {\bibinfo  {journal} {Gran. Matt.}\ }\textbf {\bibinfo {volume}
  {10}},\ \bibinfo {pages} {235–246} (\bibinfo {year} {2008})}\BibitemShut
  {NoStop}%
\bibitem [{\citenamefont {Santos}\ \emph {et~al.}(2020)\citenamefont {Santos},
  \citenamefont {Bolintineanu}, \citenamefont {Grest}, \citenamefont {Lechman},
  \citenamefont {Plimpton}, \citenamefont {Srivastava},\ and\ \citenamefont
  {Silbert}}]{santos2020}%
  \BibitemOpen
  \bibfield  {author} {\bibinfo {author} {\bibfnamefont {A.~P.}\ \bibnamefont
  {Santos}}, \bibinfo {author} {\bibfnamefont {D.~S.}\ \bibnamefont
  {Bolintineanu}}, \bibinfo {author} {\bibfnamefont {G.~S.}\ \bibnamefont
  {Grest}}, \bibinfo {author} {\bibfnamefont {J.~B.}\ \bibnamefont {Lechman}},
  \bibinfo {author} {\bibfnamefont {S.~J.}\ \bibnamefont {Plimpton}}, \bibinfo
  {author} {\bibfnamefont {I.}~\bibnamefont {Srivastava}},\ and\ \bibinfo
  {author} {\bibfnamefont {L.~E.}\ \bibnamefont {Silbert}},\ }\href
  {https://doi.org/10.1103/PhysRevE.102.032903} {\bibfield  {journal} {\bibinfo
   {journal} {Phys. Rev. E}\ }\textbf {\bibinfo {volume} {102}},\ \bibinfo
  {pages} {032903} (\bibinfo {year} {2020})}\BibitemShut {NoStop}%
\bibitem [{dis()}]{dispersity_key}%
  \BibitemOpen
  \href@noop {} {}\bibinfo {note} {We found that introducing $\pm 5\%$ diameter
  dispersity did not change the results.}\BibitemShut {Stop}%
\bibitem [{\citenamefont {Monti}\ \emph {et~al.}(2023)\citenamefont {Monti},
  \citenamefont {Srivastava}, \citenamefont {Silbert}, \citenamefont
  {Lechman},\ and\ \citenamefont {Grest}}]{monti2023}%
  \BibitemOpen
  \bibfield  {author} {\bibinfo {author} {\bibfnamefont {J.~M.}\ \bibnamefont
  {Monti}}, \bibinfo {author} {\bibfnamefont {I.}~\bibnamefont {Srivastava}},
  \bibinfo {author} {\bibfnamefont {L.~E.}\ \bibnamefont {Silbert}}, \bibinfo
  {author} {\bibfnamefont {J.~B.}\ \bibnamefont {Lechman}},\ and\ \bibinfo
  {author} {\bibfnamefont {G.~S.}\ \bibnamefont {Grest}},\ }\href
  {https://doi.org/10.1103/PhysRevE.108.L042902} {\bibfield  {journal}
  {\bibinfo  {journal} {Phys. Rev. E}\ }\textbf {\bibinfo {volume} {108}},\
  \bibinfo {pages} {L042902} (\bibinfo {year} {2023})}\BibitemShut {NoStop}%
\bibitem [{\citenamefont {Silbert}\ \emph
  {et~al.}(2002{\natexlab{a}})\citenamefont {Silbert}, \citenamefont {Grest},\
  and\ \citenamefont {Landry}}]{silbert2002-2}%
  \BibitemOpen
  \bibfield  {author} {\bibinfo {author} {\bibfnamefont {L.~E.}\ \bibnamefont
  {Silbert}}, \bibinfo {author} {\bibfnamefont {G.~S.}\ \bibnamefont {Grest}},\
  and\ \bibinfo {author} {\bibfnamefont {J.~W.}\ \bibnamefont {Landry}},\
  }\href {https://doi.org/10.1103/PhysRevE.66.061303} {\bibfield  {journal}
  {\bibinfo  {journal} {Phys. Rev. E}\ }\textbf {\bibinfo {volume} {66}},\
  \bibinfo {pages} {061303} (\bibinfo {year} {2002}{\natexlab{a}})}\BibitemShut
  {NoStop}%
\bibitem [{\citenamefont {Maher}\ \emph {et~al.}(2023)\citenamefont {Maher},
  \citenamefont {Jiao},\ and\ \citenamefont {Torquato}}]{maher2023}%
  \BibitemOpen
  \bibfield  {author} {\bibinfo {author} {\bibfnamefont {C.~E.}\ \bibnamefont
  {Maher}}, \bibinfo {author} {\bibfnamefont {Y.}~\bibnamefont {Jiao}},\ and\
  \bibinfo {author} {\bibfnamefont {S.}~\bibnamefont {Torquato}},\ }\href
  {https://doi.org/10.1103/PhysRevE.108.064602} {\bibfield  {journal} {\bibinfo
   {journal} {Phys. Rev. E}\ }\textbf {\bibinfo {volume} {108}},\ \bibinfo
  {pages} {1} (\bibinfo {year} {2023})}\BibitemShut {NoStop}%
\bibitem [{\citenamefont {Silbert}\ \emph
  {et~al.}(2002{\natexlab{b}})\citenamefont {Silbert}, \citenamefont
  {Erta\c{s}}, \citenamefont {Grest}, \citenamefont {Halsey},\ and\
  \citenamefont {Levine}}]{silbert2002-1}%
  \BibitemOpen
  \bibfield  {author} {\bibinfo {author} {\bibfnamefont {L.~E.}\ \bibnamefont
  {Silbert}}, \bibinfo {author} {\bibfnamefont {D.}~\bibnamefont {Erta\c{s}}},
  \bibinfo {author} {\bibfnamefont {G.~S.}\ \bibnamefont {Grest}}, \bibinfo
  {author} {\bibfnamefont {T.~C.}\ \bibnamefont {Halsey}},\ and\ \bibinfo
  {author} {\bibfnamefont {D.}~\bibnamefont {Levine}},\ }\href
  {https://doi.org/10.1103/PhysRevE.65.031304} {\bibfield  {journal} {\bibinfo
  {journal} {Phys. Rev. E}\ }\textbf {\bibinfo {volume} {65}},\ \bibinfo
  {pages} {031304} (\bibinfo {year} {2002}{\natexlab{b}})}\BibitemShut
  {NoStop}%
\bibitem [{\citenamefont {Stukowski}(2009)}]{ovito}%
  \BibitemOpen
  \bibfield  {author} {\bibinfo {author} {\bibfnamefont {A.}~\bibnamefont
  {Stukowski}},\ }\href {https://doi.org/10.1088/0965-0393/18/1/015012}
  {\bibfield  {journal} {\bibinfo  {journal} {Modell. Simul. Mater. Sci. Eng.}\
  }\textbf {\bibinfo {volume} {18}},\ \bibinfo {pages} {015012} (\bibinfo
  {year} {2009})}\BibitemShut {NoStop}%
\bibitem [{\citenamefont {Xie}\ \emph {et~al.}(2013)\citenamefont {Xie},
  \citenamefont {Long}, \citenamefont {Weigand}, \citenamefont {Moss},
  \citenamefont {Carvalho}, \citenamefont {Roorda}, \citenamefont {Hejna},
  \citenamefont {Torquato},\ and\ \citenamefont {Steinhardt}}]{steinhardt2013}%
  \BibitemOpen
  \bibfield  {author} {\bibinfo {author} {\bibfnamefont {R.}~\bibnamefont
  {Xie}}, \bibinfo {author} {\bibfnamefont {G.~G.}\ \bibnamefont {Long}},
  \bibinfo {author} {\bibfnamefont {S.~J.}\ \bibnamefont {Weigand}}, \bibinfo
  {author} {\bibfnamefont {S.~C.}\ \bibnamefont {Moss}}, \bibinfo {author}
  {\bibfnamefont {T.}~\bibnamefont {Carvalho}}, \bibinfo {author}
  {\bibfnamefont {S.}~\bibnamefont {Roorda}}, \bibinfo {author} {\bibfnamefont
  {M.}~\bibnamefont {Hejna}}, \bibinfo {author} {\bibfnamefont
  {S.}~\bibnamefont {Torquato}},\ and\ \bibinfo {author} {\bibfnamefont
  {P.~J.}\ \bibnamefont {Steinhardt}},\ }\href
  {https://doi.org/10.1073/pnas.1220106110} {\bibfield  {journal} {\bibinfo
  {journal} {PNAS}\ }\textbf {\bibinfo {volume} {110}},\ \bibinfo {pages}
  {13250} (\bibinfo {year} {2013})}\BibitemShut {NoStop}%
\bibitem [{vid()}]{vidoe_key}%
  \BibitemOpen
  \href@noop {} {}\bibinfo {note} {See also the video in the Supplemental
  Information.}\BibitemShut {Stop}%
\bibitem [{\citenamefont {Nowak}\ \emph {et~al.}(1997)\citenamefont {Nowak},
  \citenamefont {Knight}, \citenamefont {Povinelli}, \citenamefont {Jaeger},\
  and\ \citenamefont {Nagel}}]{nowak1997}%
  \BibitemOpen
  \bibfield  {author} {\bibinfo {author} {\bibfnamefont {E.}~\bibnamefont
  {Nowak}}, \bibinfo {author} {\bibfnamefont {J.}~\bibnamefont {Knight}},
  \bibinfo {author} {\bibfnamefont {M.}~\bibnamefont {Povinelli}}, \bibinfo
  {author} {\bibfnamefont {H.}~\bibnamefont {Jaeger}},\ and\ \bibinfo {author}
  {\bibfnamefont {S.}~\bibnamefont {Nagel}},\ }\href
  {https://doi.org/10.1016/S0032-5910(97)03291-9} {\bibfield  {journal}
  {\bibinfo  {journal} {Powd. Tech.}\ }\textbf {\bibinfo {volume} {94}},\
  \bibinfo {pages} {79} (\bibinfo {year} {1997})}\BibitemShut {NoStop}%
\bibitem [{\citenamefont {Santomaso}\ \emph {et~al.}(2003)\citenamefont
  {Santomaso}, \citenamefont {Lazzaro},\ and\ \citenamefont
  {Canu}}]{santomaso2003}%
  \BibitemOpen
  \bibfield  {author} {\bibinfo {author} {\bibfnamefont {A.}~\bibnamefont
  {Santomaso}}, \bibinfo {author} {\bibfnamefont {P.}~\bibnamefont {Lazzaro}},\
  and\ \bibinfo {author} {\bibfnamefont {P.}~\bibnamefont {Canu}},\ }\href
  {https://doi.org/10.1016/S0009-2509(03)00137-4} {\bibfield  {journal}
  {\bibinfo  {journal} {Chem. Eng. Sci.}\ }\textbf {\bibinfo {volume} {58}},\
  \bibinfo {pages} {2857} (\bibinfo {year} {2003})}\BibitemShut {NoStop}%
\bibitem [{\citenamefont {Landry}\ \emph {et~al.}(2003)\citenamefont {Landry},
  \citenamefont {Grest}, \citenamefont {Silbert},\ and\ \citenamefont
  {Plimpton}}]{landry2003}%
  \BibitemOpen
  \bibfield  {author} {\bibinfo {author} {\bibfnamefont {J.~W.}\ \bibnamefont
  {Landry}}, \bibinfo {author} {\bibfnamefont {G.~S.}\ \bibnamefont {Grest}},
  \bibinfo {author} {\bibfnamefont {L.~E.}\ \bibnamefont {Silbert}},\ and\
  \bibinfo {author} {\bibfnamefont {S.~J.}\ \bibnamefont {Plimpton}},\ }\href
  {https://doi.org/10.1103/PhysRevE.67.041303} {\bibfield  {journal} {\bibinfo
  {journal} {Phys. Rev. E}\ }\textbf {\bibinfo {volume} {67}},\ \bibinfo
  {pages} {041303} (\bibinfo {year} {2003})}\BibitemShut {NoStop}%
\bibitem [{\citenamefont {Somfai}\ \emph {et~al.}(2007)\citenamefont {Somfai},
  \citenamefont {van Hecke}, \citenamefont {Ellenbroek}, \citenamefont
  {Shundyak},\ and\ \citenamefont {van Saarloos}}]{somfai2007}%
  \BibitemOpen
  \bibfield  {author} {\bibinfo {author} {\bibfnamefont {E.}~\bibnamefont
  {Somfai}}, \bibinfo {author} {\bibfnamefont {M.}~\bibnamefont {van Hecke}},
  \bibinfo {author} {\bibfnamefont {W.~G.}\ \bibnamefont {Ellenbroek}},
  \bibinfo {author} {\bibfnamefont {K.}~\bibnamefont {Shundyak}},\ and\
  \bibinfo {author} {\bibfnamefont {W.}~\bibnamefont {van Saarloos}},\ }\href
  {https://doi.org/10.1103/PhysRevE.75.020301} {\bibfield  {journal} {\bibinfo
  {journal} {Phys. Rev. E}\ }\textbf {\bibinfo {volume} {75}},\ \bibinfo
  {pages} {020301} (\bibinfo {year} {2007})}\BibitemShut {NoStop}%
\bibitem [{\citenamefont {Song}\ \emph {et~al.}(2008)\citenamefont {Song},
  \citenamefont {Wang},\ and\ \citenamefont {Makse}}]{song2008}%
  \BibitemOpen
  \bibfield  {author} {\bibinfo {author} {\bibfnamefont {C.}~\bibnamefont
  {Song}}, \bibinfo {author} {\bibfnamefont {P.}~\bibnamefont {Wang}},\ and\
  \bibinfo {author} {\bibfnamefont {H.~A.}\ \bibnamefont {Makse}},\ }\href
  {https://doi.org/https://doi.org/10.1038/nature06981} {\bibfield  {journal}
  {\bibinfo  {journal} {Nature}\ }\textbf {\bibinfo {volume} {453}},\ \bibinfo
  {pages} {629} (\bibinfo {year} {2008})}\BibitemShut {NoStop}%
\bibitem [{\citenamefont {Silbert}(2010)}]{silbert2010}%
  \BibitemOpen
  \bibfield  {author} {\bibinfo {author} {\bibfnamefont {L.~E.}\ \bibnamefont
  {Silbert}},\ }\href {https://doi.org/10.1039/c001973a} {\bibfield  {journal}
  {\bibinfo  {journal} {Soft Matter}\ }\textbf {\bibinfo {volume} {6}},\
  \bibinfo {pages} {2918} (\bibinfo {year} {2010})}\BibitemShut {NoStop}%
\bibitem [{\citenamefont {Bililign}\ \emph {et~al.}(2019)\citenamefont
  {Bililign}, \citenamefont {Kollmer},\ and\ \citenamefont
  {Daniels}}]{bililign2019}%
  \BibitemOpen
  \bibfield  {author} {\bibinfo {author} {\bibfnamefont {E.~S.}\ \bibnamefont
  {Bililign}}, \bibinfo {author} {\bibfnamefont {J.~E.}\ \bibnamefont
  {Kollmer}},\ and\ \bibinfo {author} {\bibfnamefont {K.~E.}\ \bibnamefont
  {Daniels}},\ }\href {https://doi.org/10.1103/PhysRevLett.122.038001}
  {\bibfield  {journal} {\bibinfo  {journal} {Phys. Rev. Lett.}\ }\textbf
  {\bibinfo {volume} {122}},\ \bibinfo {pages} {038001} (\bibinfo {year}
  {2019})}\BibitemShut {NoStop}%
\bibitem [{\citenamefont {Luding}(2016)}]{luding2016}%
  \BibitemOpen
  \bibfield  {author} {\bibinfo {author} {\bibfnamefont {S.}~\bibnamefont
  {Luding}},\ }\href {https://doi.org/10.1038/nphys3680} {\bibfield  {journal}
  {\bibinfo  {journal} {Nature Phys.}\ }\textbf {\bibinfo {volume} {12}},\
  \bibinfo {pages} {531} (\bibinfo {year} {2016})}\BibitemShut {NoStop}%
\bibitem [{\citenamefont {Santos}\ \emph {et~al.}(2024)\citenamefont {Santos},
  \citenamefont {Srivastava}, \citenamefont {Silbert}, \citenamefont
  {Lechman},\ and\ \citenamefont {Grest}}]{santos2024}%
  \BibitemOpen
  \bibfield  {author} {\bibinfo {author} {\bibfnamefont {A.~P.}\ \bibnamefont
  {Santos}}, \bibinfo {author} {\bibfnamefont {I.}~\bibnamefont {Srivastava}},
  \bibinfo {author} {\bibfnamefont {L.~E.}\ \bibnamefont {Silbert}}, \bibinfo
  {author} {\bibfnamefont {J.~B.}\ \bibnamefont {Lechman}},\ and\ \bibinfo
  {author} {\bibfnamefont {G.~S.}\ \bibnamefont {Grest}},\ }\href
  {https://doi.org/10.3389/frsfm.2023.1326756} {\bibfield  {journal} {\bibinfo
  {journal} {Frontiers Soft Matter}\ }\textbf {\bibinfo {volume} {3}},\
  \bibinfo {pages} {1} (\bibinfo {year} {2024})}\BibitemShut {NoStop}%
\bibitem [{\citenamefont {Berendsen}\ \emph {et~al.}(1984)\citenamefont
  {Berendsen}, \citenamefont {Postma}, \citenamefont {van Gunsteren},
  \citenamefont {DiNola},\ and\ \citenamefont {Haak}}]{berendsen1984}%
  \BibitemOpen
  \bibfield  {author} {\bibinfo {author} {\bibfnamefont {H.~J.~C.}\
  \bibnamefont {Berendsen}}, \bibinfo {author} {\bibfnamefont {J.~P.~M.}\
  \bibnamefont {Postma}}, \bibinfo {author} {\bibfnamefont {W.~F.}\
  \bibnamefont {van Gunsteren}}, \bibinfo {author} {\bibfnamefont
  {A.}~\bibnamefont {DiNola}},\ and\ \bibinfo {author} {\bibfnamefont {J.~R.}\
  \bibnamefont {Haak}},\ }\href {https://doi.org/10.1063/1.448118} {\bibfield
  {journal} {\bibinfo  {journal} {The Journal of Chemical Physics}\ }\textbf
  {\bibinfo {volume} {81}},\ \bibinfo {pages} {3684} (\bibinfo {year}
  {1984})}\BibitemShut {NoStop}%
\bibitem [{\citenamefont {Srivastava}\ \emph {et~al.}(2021)\citenamefont
  {Srivastava}, \citenamefont {Silbert}, \citenamefont {Grest},\ and\
  \citenamefont {Lechman}}]{srivastava2021b}%
  \BibitemOpen
  \bibfield  {author} {\bibinfo {author} {\bibfnamefont {I.}~\bibnamefont
  {Srivastava}}, \bibinfo {author} {\bibfnamefont {L.~E.}\ \bibnamefont
  {Silbert}}, \bibinfo {author} {\bibfnamefont {G.~S.}\ \bibnamefont {Grest}},\
  and\ \bibinfo {author} {\bibfnamefont {J.~B.}\ \bibnamefont {Lechman}},\
  }\href {https://doi.org/10.1017/jfm.2020.811} {\bibfield  {journal} {\bibinfo
   {journal} {J. Fluid Mech}\ }\textbf {\bibinfo {volume} {907}},\ \bibinfo
  {pages} {A18} (\bibinfo {year} {2021})}\BibitemShut {NoStop}%
\end{thebibliography}%

\section{End Matter}
A physically realistic approach for generating simulated granular beds is to deposit particles through a pouring mechanism.
In the following, a weak body force due to gravity (acceleration $g = 10^{-7}k\sigma/m$) was applied to all particles, and the granular interaction parameters were identical to those used in the main text for the isotropic compression protocol.
Monodispersed particles were introduced into the deposition simulation volume iteratively and fell from rest approximately $10\sigma$ above the bed surface---exploration of effects pertaining to air drag and the kinetic energy and incidence angle of deposited particles is left to future work.  
A total of two million particles were deposited in each simulation into a domain bounded by a flat, frictionless wall located at $z = 0$.
The lateral dimensions of the domain were $160\sigma \times 160\sigma$ and periodic boundary conditions in the plane of the bounding wall were used.
The sliding friction coefficient, $\mu_s$, acting between particles was varied from $0-0.3$ across four simulations ($\mu_r = \mu_t = 0$), with all other parameters held constant.
The typical bed thickness produced by deposition under these conditions was greater than $60\sigma$.

\begin{figure*}
\centering
\includegraphics[width=\textwidth]{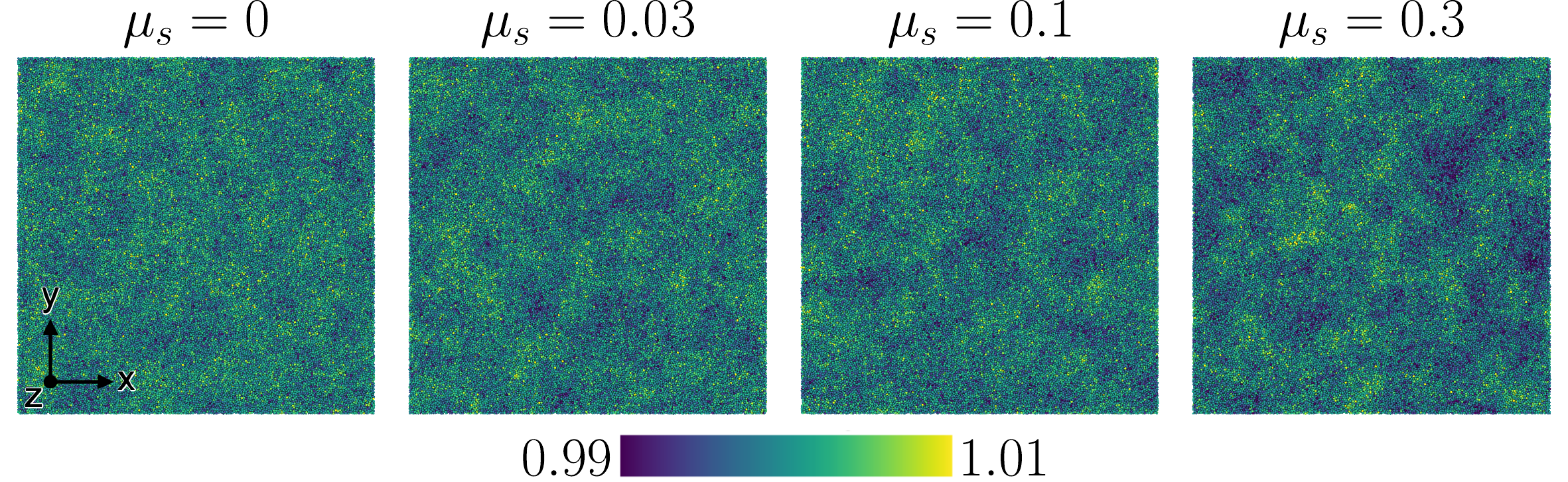}
\caption{Top-down images of poured granular beds showing internal coordination variations for the indicated $\mu_s$ and $W = 16$. 
The displayed vertical slices span $48\sigma \leq z \leq 52\sigma$ (for bed depths greater than $60\sigma$) and have lateral dimensions $160\sigma \times 160\sigma$.}
\label{fig:figEM}
\end{figure*}

Deposition onto a featureless wall produces layering in the immediate vicinity of the wall---up to $z \approx 10 \sigma$ for small $\mu_s$.
Moreover, without bounding frictional walls to support the weight of the bed (via the Janssen effect), density gradients extending through the bed thickness also formed depending on $\mu_s$.
To isolate coordination heterogeneity within the bed, for which lateral heterogeneity provides the best analogy to the results in the main text, Fig.~\ref{fig:figEM} shows top-down views of vertical slices through the deposited beds ($48\sigma \leq z \leq 52\sigma$), colored by coordination variations obtained using a spherical window of diameter $16\sigma$. 
Mimicking the depictions in Fig. 2, particle coordination is normalized by $\phi_s W^3$, where $\phi_s$ is the packing density of the $4\sigma$-wide vertical slices: $\phi_s = 0.641$ for $\mu_s = 0$, $\phi_s = 0.637$ for $\mu_s = 0.03$, $\phi_s = 0.630$ for $\mu_s = 0.1$, and $\phi_s = 0.616$ for $\mu_s = 0.3$. 

Figure~\ref{fig:figEM} shows that increasing $\mu_s$ produces an associated increase in perceptible spatial variations of particle coordination over length scales comparable to $16\sigma$.
The structure indicated in Fig.~\ref{fig:figEM} for $\mu_s = 0.1 - 0.3$ is weaker in magnitude than was observed for the corresponding isotropically compressed packings described in the main text.
2D $S(q)$ calculations performed on the deposited particle beds (calculated requiring $q_z = 0$) did not exhibit a notable prepeak, in contrast to $S(q)$ obtained for the  isotropic packings.
Despite this difference, the visible signature of intermediate range order in Fig.~\ref{fig:figEM} follows the same trend of increasing order with increasing friction, reinforcing the notion that strong frictional interactions are capable of stabilizing heterogeneous structures over length scales much larger than the particle size, while granular assemblies of frictionless particles generally lack structure on such large length scales.

\end{document}